\documentclass[10pt,a4paper,twoside]{article}
\usepackage{epsfig}
\usepackage{baltlat5}
\begin{document}
\ \
\vspace{-0.5mm}

\setcounter{page}{1}
\vspace{-2mm}

\titlehead{Baltic Astronomy, vol.\ts xx, xxx--xxx, 2008.}

\titleb{CHEMICAL EVOLUTION OF THE GALACTIC BULGE: \\ SINGLE AND DOUBLE INFALL MODELS}

\begin{authorl}
\authorb{R. D. D. Costa}{}
\authorb{W. J. Maciel}{} and
\authorb{A. V. Escudero}{}
\end{authorl}

\begin{addressl}
\addressb{}{University of S\~ao Paulo, Rua do Mat\~ao 1226, Cidade Universit\'aria, 
04005-000 S\~ao Paulo SP, Brazil}
\end{addressl}


%

\submitb{Received 2008; revised 2008}

\begin{summary}
Recent work has produced a wealth of data concerning the chemical evolution of the galactic
bulge, both for stars and nebulae. Present theoretical models generally adopt a limited
range of such constraints, frequently using a single chemical element (usually iron), which
is not enough to describe it unambiguously. In this work, we take into account constraints
involving as many chemical elements as possible, basically obtained from bulge nebulae
and stars. Our main goal is to show that different scenarios can describe, at least partially, 
the abundance distribution and several distance-independent correlationss for these objects.  
Three classes of models were developed. The first is a one-zone, single-infall model, 
the second is a one-zone, double-infall model and the third is a multizone, double infall
model. We show that a one-zone model with a single infall episode is able to reproduce 
some of the observational data, but the best results are achieved using a multizone, 
double infall model. 
\end{summary}



\resthead{The galactic bulge: single and double infall models}
{R.D.D. Costa, W. J. Maciel, A. V. Escudero}

\sectionb{1}{INTRODUCTION}

The galactic bulge has been extensively studied in the last few years, but many of its
properties and formation history are still open to discussion. Among the main bulge
characteristics that can be taken as constraints for chemical evolution models are
the metallicity distribution, the $\alpha$-element relation to the metallicity, and 
several abundance correlations that are distance-independent. Concerning stellar data,
such constraints have had a considerable improvement in the last couple of years
(see for example Rich and Origlia 2005, Cunha and Smith 2006, Fulbright et al. 
2006, 2007, Rich et al. 2007, Zoccali et al. 2006, and Lecureur et al. 2007).
On the other hand, nebular data have also improved, as can be seen from our own results
(Cuisinier et al. 2000, Escudero and Costa 2001, Escudero et al. 2004, Cavichia et
al. 2008). Although the chemical abundances of planetary nebulae (PNe) can be obtained with
a high accuracy for several elements that are more difficult to study in stars,
these results are often overlooked in the literature, despite their importance as
constraints for chemical evolution models of the bulge.

Regarding the bulge formation and evolution, a mixed scenario seems to be more attractive. 
Since the earlier models for bulge stars that could predict ratios of $\alpha$-elements 
for bulge metallicities (Matteucci and Brocato 1990), several models adopting a single 
fast collapse have been proposed. These models are able to explain isolated stellar 
abundances, but the abundance correlations for large groups of chemical elements are 
generally not well reproduced. Besides, kinematic evidences point to a bulge rotation 
profile similar to a disk, but including an additional component with a larger velocity 
dispersion (Beaulieu et al. 2000). Some recent models have at least partially corrected
this situation, as in the theoretical models by Ballero et al. (2006, 2007a,b). In 
Ballero et al. (2007b), to which the reader is referred for a detailed discussion
on previous chemical evolution models for the galactic bulge, observational constraints
such as the metallicity distribution and $\alpha$-element ratios as a function
of metallicity are well reproduced, especially for oxygen, for which a large variety of
observational data is available. The authors claim that there is no need to invoke a 
second infall episode, but it should be noted that the abundance correlations taken as 
constraints are limited. In fact, it has become increasingly more difficult to explain 
all these observations in a satisfactory way using a single infall episode.  

In this work we present single- and double-infall models for the bulge evolution, using
both planetary nebulae and stellar data as constraints. The PNe data comes basically from
our own group, while for stars we have used recent data from the literature. In order
to describe the galactic bulge in a less ambiguous way, we made an effort to include as 
many chemical elements as possible. Our main goal is to show that different scenarios 
can describe, at least partially, the abundance distribution and other abundance correlations
for bulge objects. We show that a one-zone model with a single infall episode is able to 
reproduce some of the abundance distributions, but the best results are achieved using 
a multizone, double infall model.

\sectionb{2}{THE CHEMICAL EVOLUTION MODEL}

\subsectionb{2.1}{The Star Formation Rate}
The star formation rate (SFR) is a key factor to describe the chemical evolution of a  
galaxy, as it gives the total amount of gas converted into stars, which depends on many 
environmental factors, such as density, temperature, presence of winds, tidal forces, etc.
The SFR affects directly nearly all the results of a chemical evolution model, since it 
modifies not only the amount of stars, but also the gas density of the medium. Most 
approximations for the SFR are power laws of the gas density, and the usual form is given 
by the generic Schmidt law:

   \begin{equation}
       SFR = c \ \sigma^{k},
   \end{equation}
   
\noindent 
where $c$ is a constant, $\sigma$ is the gas density and $k$ a constant greater than unity.
The values of the constants $c$ and $k$ are usually derived empirically from observations of 
spirals and starburst galaxies (Schmidt 1959, Buat et al. 1989, Kennicutt 1998a,b). We have
adopted the values derived by Kennicutt (1998b), $c = (2.5\pm 0.7)\times 10^{-4}$ and 
$k = 1.4\pm 0.15$, so that the SFR is given in $M_\odot\,{\rm year^{-1}}\, {\rm kpc^{-2}}$. 
In Equation (1), $\sigma$ is then the surface density, which can be related to the
average volume density of the gas. These values are generally similar to the values based
on SFR derived from H$_{\alpha}$, UV and FIR data, as given by Buat et al. (1989), 
Kennicutt (1989), Buat (1992), Boselli (1994), Deharveng et al. (1994), and Boselli et al. 
(1995).

\subsectionb{2.2}{Infall}
The infall rate is not a well known parameter, and depends on many factors such as the 
total amount of mass, gas density and gas collisions. We can define the infall rate as
an exponential profile of the form $\dot M(t) \propto \exp(-t/\tau)$, where $\tau$ is 
the input timescale to the medium (Chiosi 1980). In spite of being a simplified way to 
represent the gas increase rate, chemical evolution models have shown good agreement 
when compared to observational constraints (Chiappini et al. 1997). We can then write

    \begin{equation}
        \dot M(t) = A\ e^{-t/\tau} \ .
    \end{equation}

The proportionality constant can be estimated from the total amount of material, $M_T$:

    \begin{equation}
        M _{T} = A \tau \ .
    \end{equation}

To obtain the amount of material falling in a given time interval, the infall rate has to be 
integrated as:

    \begin{equation}
        \Delta M = \int_{t}^{t + \Delta t}{\dot M(t) dt} \ .
    \end{equation}

\noindent 
Therefore, we have an expression for the amount of material accreted by the system for any 
time interval. However, in an open system such the Galaxy, where the halo mass can be 
constantly altered due to winds, accretion of satellites or other mechanisms, it is 
convenient to express this value as a function of the remaining infall mass:

    \begin{equation}
        M _F = \int_{t}^{\infty}{\dot M(t) dt}
    \end{equation}

\noindent
Combining equations (4) and (5) we have:

    \begin{equation}
        \Delta M(t) = M _F \left(1 - e^{ -\frac{\Delta t}{\tau} } \right)      
    \end{equation}

\subsectionb{2.3}{Binary Systems and SNIa}
Stars in binary systems can have a different evolutionary path compared to individual stars. 
Depending on the mass and separation of the components in a pair they can evolve to a 
Type-Ia supernova when one of the components reaches the Chandrasekhar limit. The way 
stars combine depends on their mass ratio and lifetime of the components, so that the 
number of SNIa for each mass interval of the secondary may affect the abundances of the 
medium. In this work we adopted the formalism by Ferrini et al. (1992, see also Matteucci
and Greggio 1986). More recent tratments (cf. Matteucci et al. 2006) based on the
assumption that Type-Ia SN originate from CO white dwarfs in binary systems without
specifying the degeneracy of the progenitor stars may affect the earlier star formation
epochs of the galactic halo, but the main conclusions of this paper for the bulge
evolution are probably unchanged. 

\vskip 1.0 true cm

\subsectionb{2.4}{The IMF and stellar yields}
Stars of different masses have different lifetimes and chemical yields. The IMF is a 
necessary ingredient of any  chemical evolution model, as it gives essentially the 
total amount of stars in each mass interval, which basically affects the stellar yields.
In this work we used the Kroupa (2002) IMF, which provides a more realistic distribution 
of objects with respect to the observational data. For the initial period of multizone, 
double infall models, we also used Salpeter's IMF, as explained in Section 4.2. 
Metallicity dependent yields are still quite uncertain (see for example Matteucci 2001, 
McWilliam et al. 2008), so that most applications adopt mass-dependent yields. In this work, 
we have used stellar yields derived from numerical models by van den Hoek \& Groenewegen 
(1997)  for intermediate mass stars and Tsujimoto et al. (1995) for Type-II supernovae 
and SNIa.

\subsectionb{2.5}{Winds}
Galactic winds are important in many chemical evolution models in order to explain the 
observed abundances in dwarf and elliptical galaxies. However, a detailed and self 
consistent treatment of the mass loss is too complex, as it depends on factors such as
the presence of infall, enviromental gas, external pressures, geometrical distribution,
etc. Therefore, the mass  loss in the galactic bulge is usually included as a free
parameter, as in Ferreras et al. (2003). In this work, we have adopted a similar 
procedure, and the wind effect was simulated by considering that a fraction of the 
material ejected by SNII/Ia, considered as a free parameter, is lost to an adjacent 
region, to the halo or out of the Galaxy.

\sectionb{3}{THE OBSERVATIONAL SAMPLES}

Concerning planetary nebulae, the observations and data reduction procedures are 
described by Escudero et al. (2004), to which the reader is referred for details. 
Observations were performed in two telescopes: 1.60 m LNA (Bras\'opolis - Brasil) 
and 1.52 ESO (La Silla – Chile) from 2001 to 2003. In both observatories, the observations 
consisted in long slit spectroscopy using Cassegrain spectrographs, with gratings of 
300 l/mm and 600 l/mm respectively, resulting in reciprocal dispersions of 4.4 \AA /pixel 
and 2.2 \AA /pixel. Some additional data on PNe taken from the literature are also 
described in the same paper. The PNe sample is contained within about $\pm 7$ degrees in 
galactic latitude, to make sure that bulge nebulae only are included. The region within 
about $\pm 1$ degree is underpopulated in all PNe samples, in view of the large 
extinction in this region.

Regarding stellar data, several recent sets of observations of bulge stars have also 
been taken into account. These comprise basically the results of Zoccali et al. (2003, 
2006, 2008), Rich and Origlia (2005), Cunha and Smith (2006), Fulbright et al. (2006, 2007), 
Rich et al. (2007) and Lecureur et al. (2007). For the sake of completeness, the set 
of bulge-like stars by Pomp\'eia et al. (2003) has also been taken into account. Since 
these objects may be inner disk stars, a comparison of the results from this sample and 
the remaining ones may shed some light on the apparently different evolution of the bulge 
and inner disk.

\vspace{10mm}

\sectionb{4}{MODELS}

\subsectionb{4.1}{One-zone models}
Two kinds of one-zone models were elaborated: single and double infall models. Table~1
displays the input parameters adopted for these models. The parameters in columns 2 and
3 are {\it not} free parameters, and were derived independently of the chemical evolution: 
M1, the mass of the first infall, corresponds to the mass of the spheroidal component as 
derived by Amaral et al. (1996); M2, the mass of the second infall, corresponds to the mass 
of the discoidal component, within the first 1.5 kpc, and was calculated based on the disk 
radial density profile. Regarding the infall timescales ($\tau_1, \tau_2$) and the wind rate 
$W$, estimates of the parameter space were made on the basis of model calculations by Moll\'a 
et al. (2000). The results for $\tau_1, \tau_2, $ and $W$ are shown in columns 4--6 of 
Table~1, respectively.

\vspace{3mm}

\begin{center}
\vbox{\small
\tabcolsep=15pt
\begin{tabular}{lccccc}
\multicolumn{6}{c}{\parbox{90mm}{
{\normbf \ \ Table 1.}{\norm\ Parameters for the one-zone models.}}}\\
\tablerule
\multicolumn{1}{c}{Model}&
\multicolumn{1}{c}{M1}&
\multicolumn{1}{c}{M2}&
\multicolumn{1}{c}{$\tau$1}&
\multicolumn{1}{c}{$\tau$2}&
\multicolumn{1}{c}{W}\\
\multicolumn{1}{c}{}&
\multicolumn{1}{c}{(M$_{\odot}$)}&
\multicolumn{1}{c}{(M$_{\odot}$)}&
\multicolumn{1}{c}{(Gyr)}&
\multicolumn{1}{c}{(Gyr)}&
\multicolumn{1}{c}{}\\
\tablerule
    Single & 1.24E10 & - & 1.0  & - & 40\% \\
    Double & 1.24E10 & 2.26E9 & 0.1 & 2.0 & 60\% \\
\tablerule
\end{tabular}
}
\end{center}
\vspace{-2mm}

\subsectionb{4.2}{Multizone double-infall model}
The multizone model is based on a mixed scenario for the bulge evolution. To reproduce the 
abundance distribution and the correlations between elemental abundances, the adopted model 
has two main phases: the first one is a fast collapse of the primordial gas, essentially 
responsible for the bulge formation, and the second is a slower infall of enriched gas that 
forms the disk. The bulge and central region of the Galaxy were divided into two zones (cf.
Table~2), the first one experiencing two gas infall episodes, a 0.1 Gyr collapse and an 
enriched gas infall lasting 2.0 Gyr, as in the previous models of Table~1. Such a division 
into two zones was chosen as it reproduces in a more realistic way the galaxy evolution 
scenario of a first infall forming the central region (zone-0 in the present model) and 
a second one, of enriched gas, to form the disk. The zones have been divided into concentric 
rings of radius $R1$, $R2$ at 1.5 kpc intervals, as shown in Table~2. The infall masses 
are also given in the table, where the second infall mass was derived from the disk 
density (Rana 1991). 

\vspace{3mm}

\begin{center}
\vbox{\small
\tabcolsep=15pt
\begin{tabular}{lcccc}
\multicolumn{5}{c}{\parbox{90mm}{
{\normbf \ \ Table 2.}{\norm\ Parameters of the multizone model.}}}\\
\tablerule
\multicolumn{1}{c}{zone}&
\multicolumn{1}{c}{R1 (kpc)}&
\multicolumn{1}{c}{R2 (kpc)}&
\multicolumn{1}{c}{M1(M$_{\odot}$)}&
\multicolumn{1}{c}{M2(M$_{\odot}$)}\\
\tablerule
    0 & 0.0 & 1.5  & 1.24E10 & 2.26E9 \\
    1 & 1.5 & 3.0  & 0 & 4.75E9 \\
\tablerule
\end{tabular}
}
\end{center}
\vspace{2mm}

To better reproduce the chemical abundances for low mass objects, the bulge evolution was 
divided into two periods with a different IMF. The duration of each period was selected 
in order to achieve a best fit when comparing model predictions to the observational data. 
Based on hydrodynamical simulations (Samland et al. 1997), we assumed that, at the 
beginning of the bulge formation, a large amount of the elements produced by SNe can 
be ejected to the halo and inner disk. Investigations on the variation of the slope of the
IMF with respect to physical parameters of the ISM show that it can depend both on the 
metallicity (Silk 1995) and on temperature and density (Padoan et al. 1997). At the 
beginning of the bulge formation, the bulge ISM was dense and probably had high velocity 
dispersion. Adopting this hypothesis, Salpeter's IMF was used in the first 0.6 Gyr, and 
then Kroupa's IMF for the rest of the evolution. For the first period, we adopted Salpeter's 
IMF, assuming that 85\% of the elements produced by SNe are ejected for the halo and inner 
disk,  70\% of which are ejected out of the bulge and 15\% from zone $0$ to zone $1$. 
For the subsequent evolution we adopted the Kroupa (2002) IMF with a wind rate of 60\%, 
divided in 45\% of which are ejected to outside the bulge and 15\% from zone $0$ to zone $1$. 
It should be noted that, since zone $1$ is formed only by the second gas infall, it does not 
experience the first period effects.

\sectionb{5} {RESULTS}

\subsectionb{5.1}{Abundance distributions of $\alpha$-elements}

Figure~1 displays the chemical abundance distributions of O, Ar, S and Ne for bulge PNe 
(histograms) compared to the model predictions.  The abcissae show the abundances by number 
of atoms of each element X, defined as $\epsilon({\rm X}) = \log ({\rm X/H}) + 12$, and the 
ordinate gives the fraction of objects in the whole sample. In this figure, red lines 
represent one-zone models: continuous red lines for single infall models and dashed red 
lines for models with double infall. Black lines represent two-zone models as follows: 
black continuous lines for the central region (zone 0) and black dashed lines for the 
outer region (zone 1).

It can be seen that the observed oxygen abundance distribution shows a good agreement 
with both classes of models. Oxygen abundances in PNe reflect the interstellar abundances 
at the time the progenitor stars were formed, although a small depletion may be observed 
due to ON cycling for the more massive progenitors. Results for one-zone and multizone 
models do not differ strongly, except that zone 1 of the multizone model is more metal 
rich than zone 0, as expected. A similar effect is apparent for argon and neon.
The agreement of the model and observational data in Figure~1 is generally reasonable,
considering the simplicity of the models and the fact that the data samples are incomplete.
The largest discrepancies between the models and the observed distributions occur for
neon, in which case the observed abundance distribution peaks about 0.2 dex higher than
predicted. This is typically the expected uncertainty of the data, which can explain
the discrepancy, along with the simplicity of the model, as mentioned.

The argon and sulfur data produce a generally better agreement with the models. In the 
one-zone, one infall model, an infall timescale of 1.0 Gyr was adopted, which makes most 
of the SNII to be produced in a short time interval. As a  consequence, oxygen and neon 
are also produced in a short time interval, rapidly enriching the ISM. Argon and sulfur, 
on the other hand, are also produced by SNIa, taking a longer time to be ejected to the 
ISM with respect to SNII yields. This effect can be seen in the figure, where the solid 
red lines show a smaller fraction of enriched material in S and Ar as compared to the 
dashed red line. 

The small differences between model results for the abundance distributions can be an 
indication that, in the case of PNe, we are observing preferentially objects coming from 
the first collapse. Most of the objects in the sample are found in relatively high 
latitudes, which supports this hypothesis.
	
\vspace{5mm}
\vbox{
\centerline{\psfig{figure=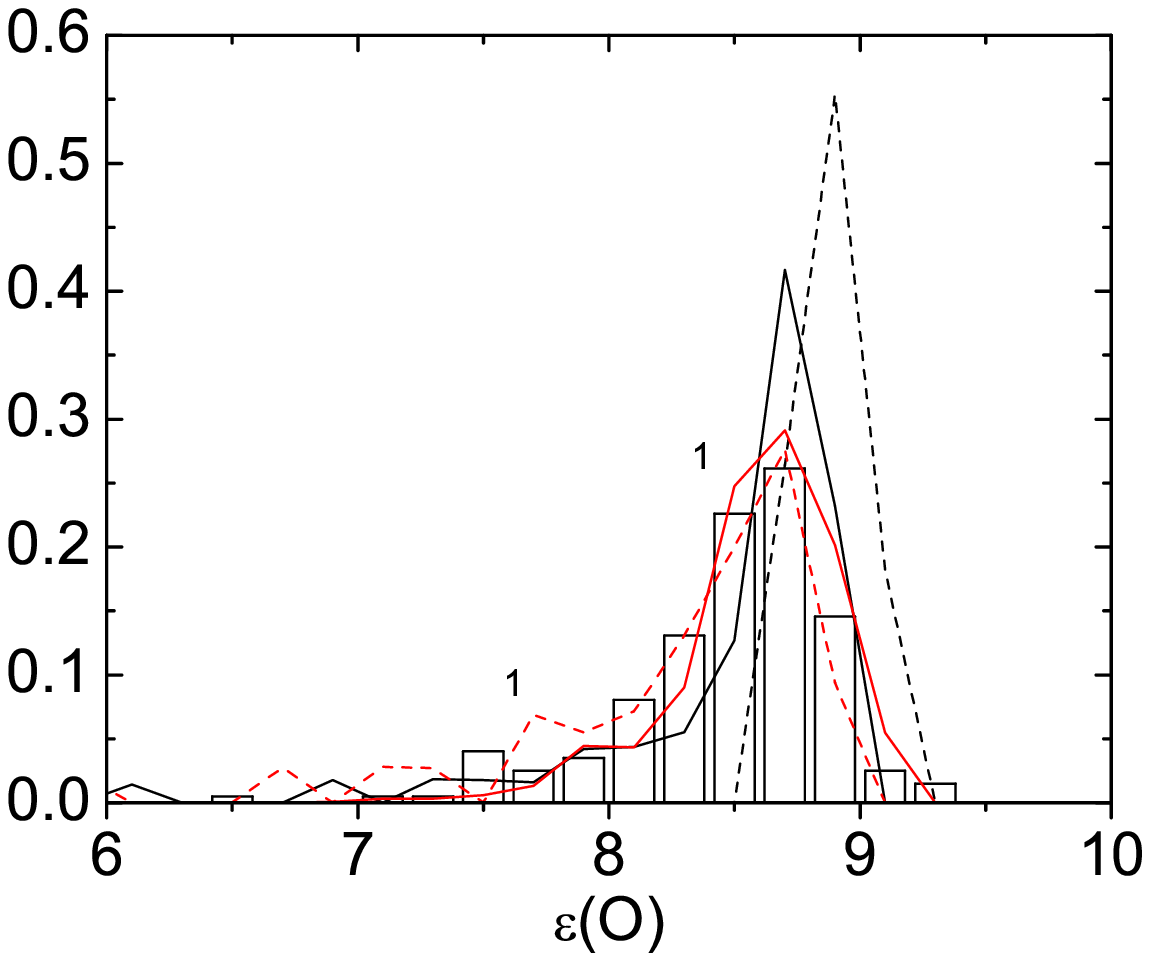,width=65truemm,angle=0,clip=}
\psfig{figure=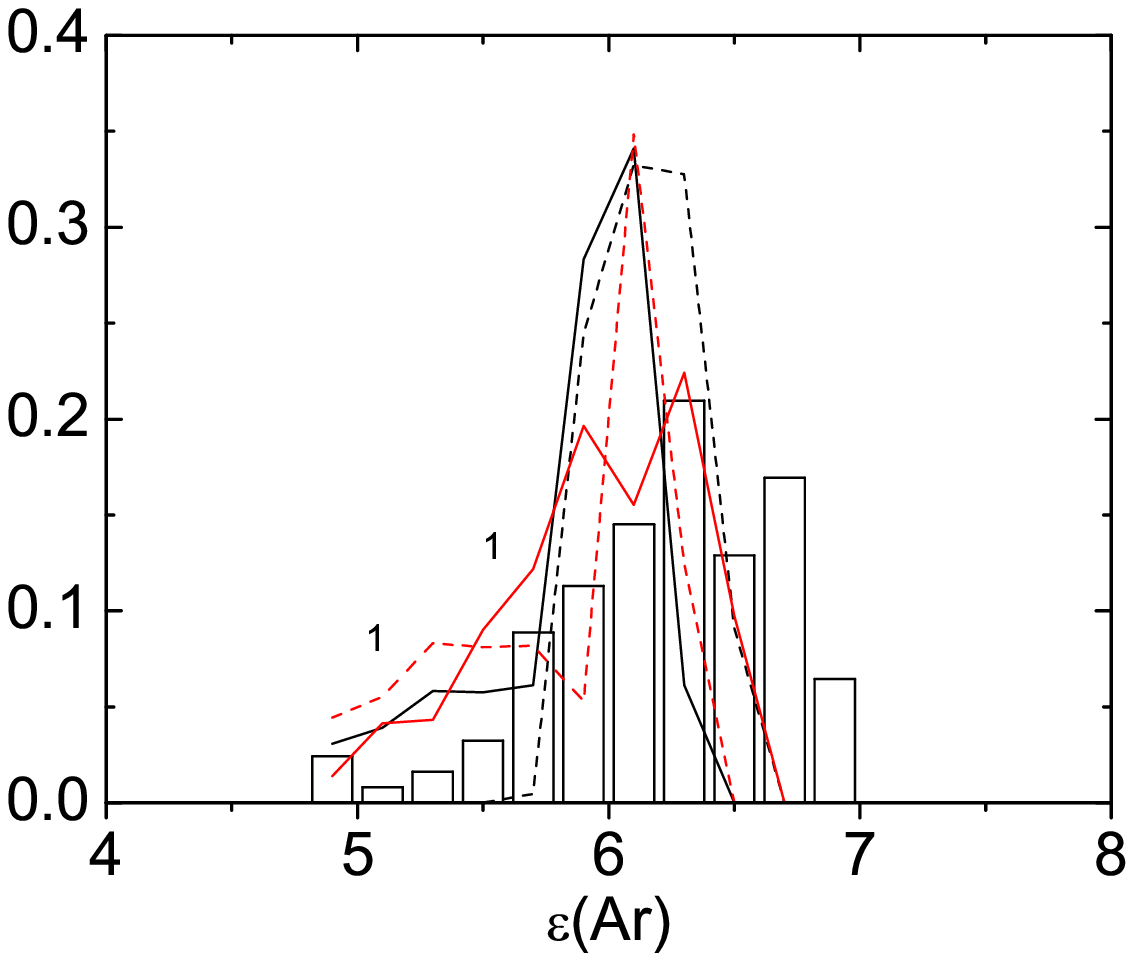,width=65truemm,angle=0,clip=}}
\centerline{\psfig{figure=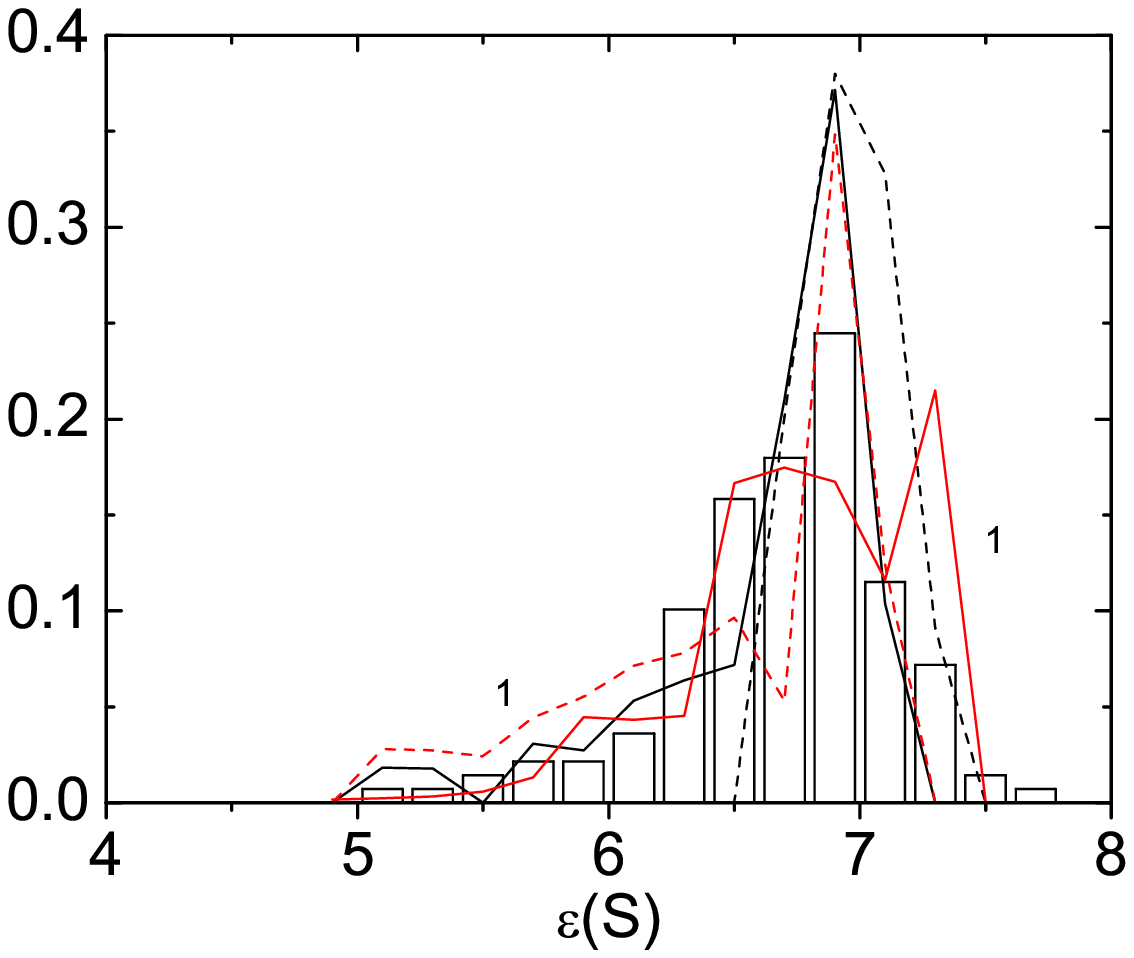,width=65truemm,angle=0,clip=}
\psfig{figure=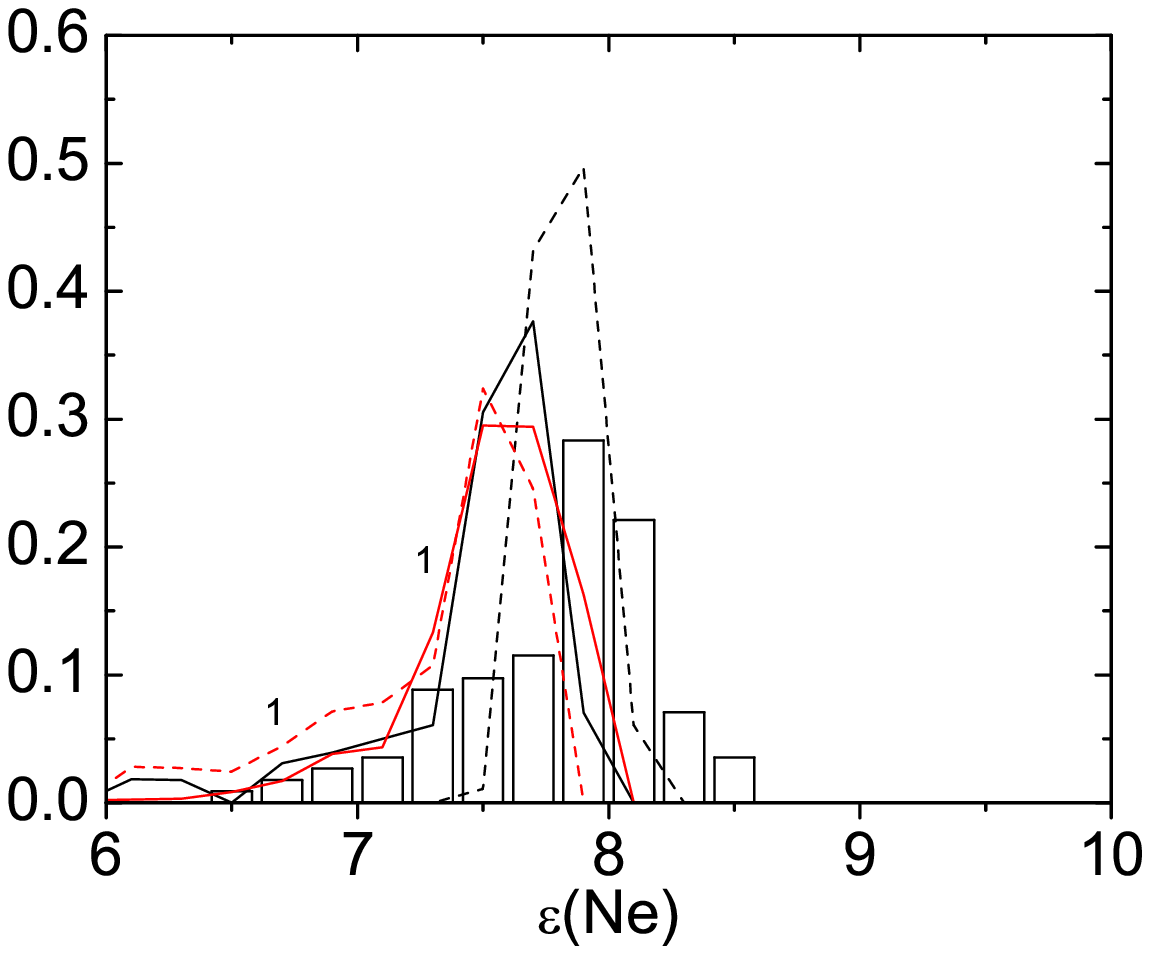,width=65truemm,angle=0,clip=}}
\vspace{-5mm}
\captionc{1}{Abundance distribution of $\alpha$-elements derived from PNe (histograms) 
       compared to model predictions. The abcissae show the elemental abundances 
       $\epsilon$(X) = log X/H + 12. Results for one-zone models are shown as red lines: 
       single infall (red continuous lines) and double infall (red dashed lines). For 
       clarity, these models are labelled as \lq\lq 1\rq\rq. Results 
       for multizone  models are shown in black: central region (region 0, black continuous 
       lines) and outer region (region 1, black dashed lines).}}
\vspace{5mm}

From Figure~1 we already have an indication that the one-zone models considered in this 
paper are not able to completely reproduce the observational data. Although a general 
agreement is achieved, the distributions generally do not match all the observations. 
In particular, the need of more complex models is apparent from the neon data 
shown in Figure~1, where the multizone model produces an improved agreement. Such a need 
is reinforced when one tries to increase the observational constraints by analyzing 
distance independent abundance correlations, as we will see in the following sections.

\vspace{5mm}

\subsectionb{5.2}{Metallicity distribution in the bulge}

Another important constraint of chemical evolution models for the bulge is the observed 
metallicity distribution of bulge stars, as measured by the [Fe/H] ratio. Iron abundances 
cannot be accurately derived from PNe, since the corresponding lines are very weak and the 
abundances eventually derived are hampered by the fact that a significant fraction of this 
element is locked up in grains (cf. Perinotto et al. 1999). Therefore, our predicted 
metallicity distribution for the bulge, as shown in Figure~2, should be compared with recent 
determinations of the [Fe/H] metallicity distribution in bulge stars. Such a determination 
has been provided by Zoccali et al. (2003), based on a combination of near-IR data with 
optical data. As shown by the histogram of Figure~2, the Zoccali et al. (2003) distribution 
peaks at near solar value, with a sharp cutoff just above the solar metallicity, and 
presenting a tail towards lower metallicities down to approximately $-$1.5 dex. These 
characteristics are generally well reproduced by our models, particularly in the case of 
the multizone model. 

\vspace{5mm}
\vbox{
\centerline{\psfig{figure=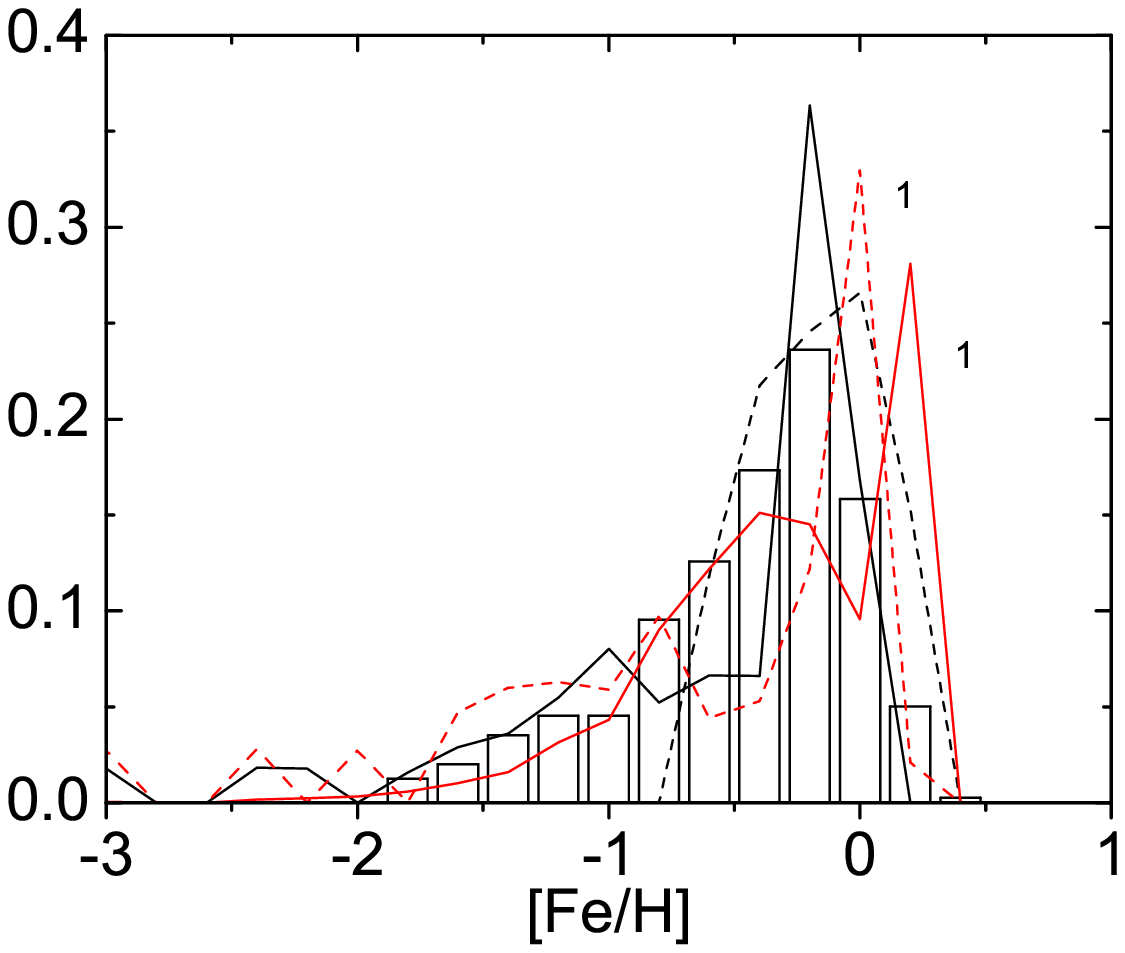,width=65truemm,angle=0,clip=}
\psfig{figure=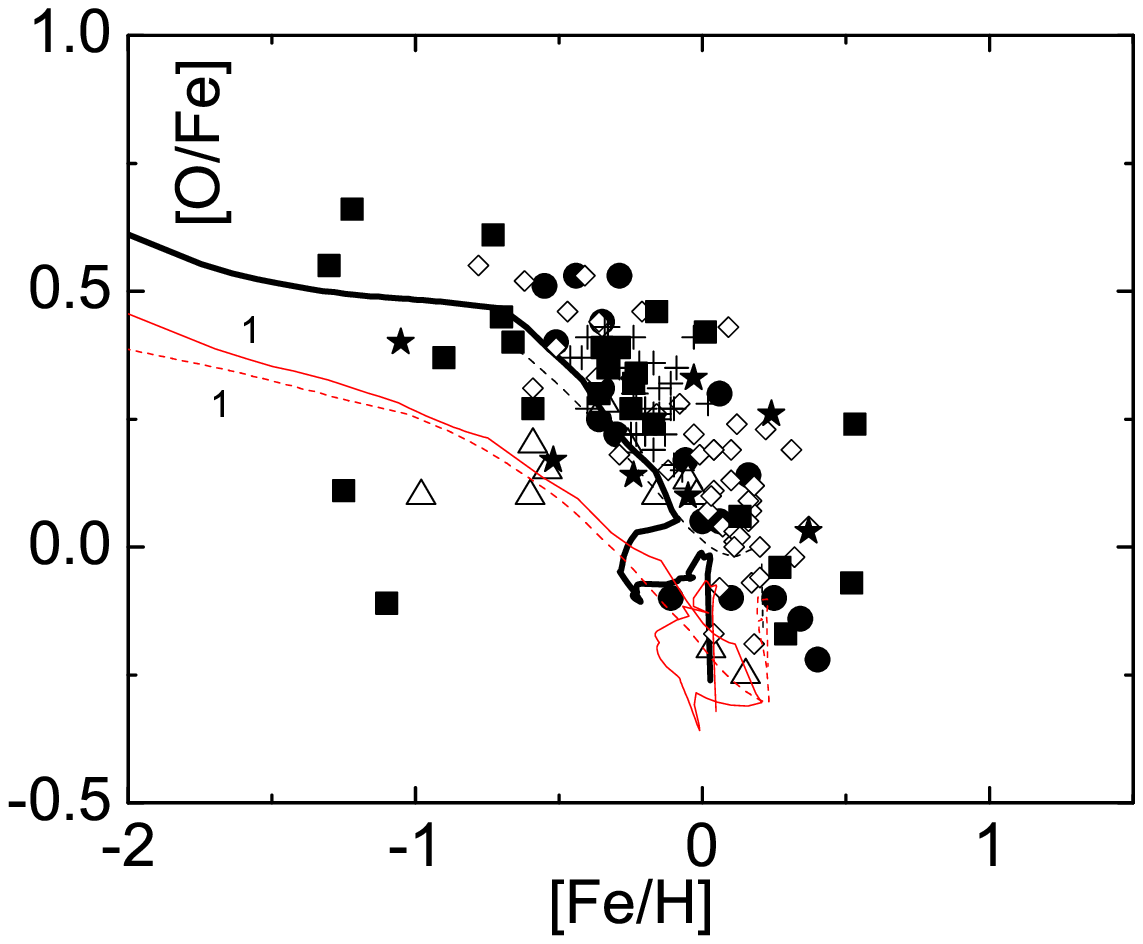,width=65truemm,angle=0,clip=}}
\vspace{-5mm}
\captionc{2}{(Left) Metallicity distribution of bulge stars (Zoccali et al. 2003, histogram) 
             and model predictions. Symbols are the same as in Figure~1.}
\captionc{3}{(Right) The [O/Fe] $\times$ [Fe/H] relation from our models compared to 
        observational data (see text).  Model symbols are the same of Figure~1.}}
\vspace{5mm}

More recently, the bulge metallicity distribution was determined by Rich and Origlia (2005) 
and Rich et al. (2007) based on infrared spectroscopy of M giants in Baade's Window and in 
the inner bulge. Although the samples are relatively small, they both agree in the sense 
that the distribution peaks about [Fe/H] $\simeq\,-0.2$, in good agreement with the 
prediction of our multizone model shown in Figure~2. The metallicity distribution of K giants 
in Baade's Window has also been derived by Fulbright et al. (2006) based on Keck HIRES 
spectra, with the result that the metallicities range approximately from [Fe/H] 
$\simeq\,-1.3$ to +0.5, peaking around [Fe/H] $\simeq\,-0.10$, again near solar value, in 
agreement with the results previously mentioned. Also, a new recalibration of previous data 
shows essentially the same characteristics, namely a peak near solar value and a tail 
towards lower metallicities (see Fulbright et al. 2006 for details). Finally, Zoccali et al.
(2008) have presented a detailed metallicity distribution in three fields of the galactic
bulge based on 800 K giants using for the first time high resolution spectroscopy of 
individual stars. The iron distribution function (IDF) for the field $b = -6^{\rm o}$
is roughly similar to the photometric distribution of Zoccali et al. (2003) which is shown
in Fig.~2, but extends to slightly larger metallicities, by about 0.2 dex, and presents
an excess of objects with metallicities at both sides around -1 dex. As can be seen
from Fig.~2, our models have just the same characteristics, extending to slightly higher
metallicities than the previous data of Zoccali et al. (2003), and in fact presenting
some excess around -1 dex. Therefore, we may conclude that our models, especially the multizone 
models, show a quite reasonable agreement with recently derived metallicity distributions 
of bulge stars.

It is interesting to compare the metallicity distribution of Fig.~2 with data from
bulge-like stars by Pomp\'eia et al. (2003). The metallicity distribution of these stars 
is similar to the distribution of true bulge stars, in the sense that it peaks slightly 
below solar and shows a tail at lower metallicities. This supports the conclusion by Zoccali 
et al. (2003) and Fulbright et al. (2006) that there is no evidence of any major abundance 
gradient in the inner bulge, but it should be noted that Zoccali et al. (2008) found some 
indication of a small gradient in larger fields, and suggested a double-component structure 
comprising an inner pseudo-bulge within an outer classical bulge.

\subsectionb{5.3}{[$\alpha$/Fe] correlations with metallicity}
The metallicity dependence of the [O/Fe] ratio of bulge stars is possibly the single 
most important constraint of chemical evolution models regarding abundance correlations.
Figure~3 compares our model predictions for this relation in the galactic bulge
with data for bulge stars taken from different sources: Barbuy \& Grenon (1990)
[triangles], Cunha \& Smith (2006) [stars], Fulbright et al. (2007) [squares], Lecureur et 
al. (2007) [diamonds], Rich \& Origlia (2005) and Rich et al. (2007) [crosses]. Model 
symbols are the same as in Figure~1. These objects are believed to define the behaviour 
of the [O/Fe] $\times$ [Fe/H] relation in the galactic bulge with a reasonable accuracy, 
as they include recent, accurate abundance determinations. 

It can be seen that the multizone model shows a good fit to the observational data, with
a clear improvement relative to the one-zone models. In this case, the difference between 
our two types of models is much larger than, for instance, in the case of the abundance 
distribution of PNe data shown in Figure~1. This is a strong indication that a more 
complex model for the bulge evolution is required. Another feature of our models that 
is supported by the observations is the relatively low [O/Fe] ratio predicted at low 
metallicities, in contrast with the models by Ballero et al. (2007b), in which a value 
of [O/Fe] $\simeq 1$ is obtained. In Figure~3 we also plot the sample of bulge-like stars 
by Pomp\'eia et al. (2003) [circles]. As in the case of the metallicity distribution 
(Figure~2), there is no appreciable differences between this sample and the remaining 
ones, which consist of true bulge stars.

\vspace{5mm}
\vbox{
\centerline{\psfig{figure=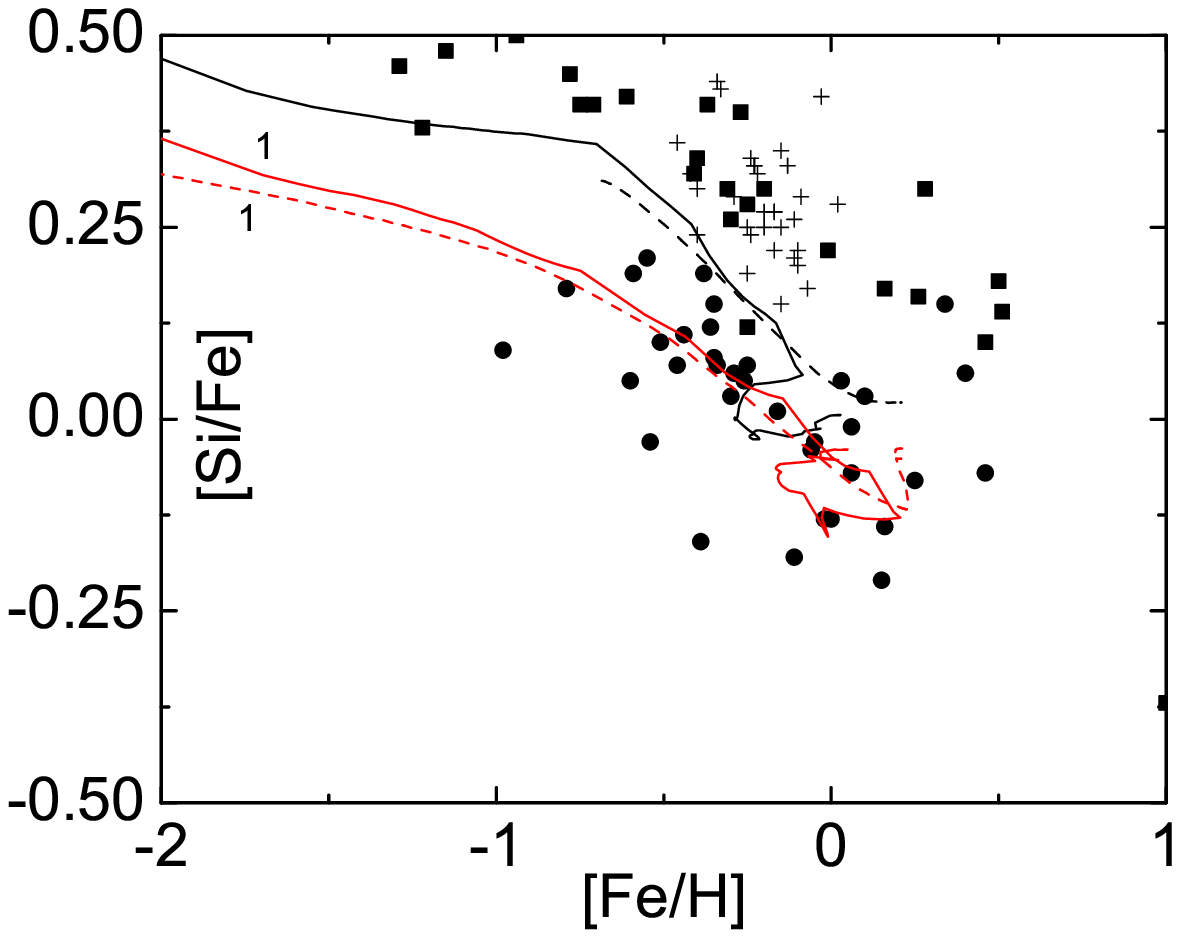,width=65truemm,angle=0,clip=}
\psfig{figure=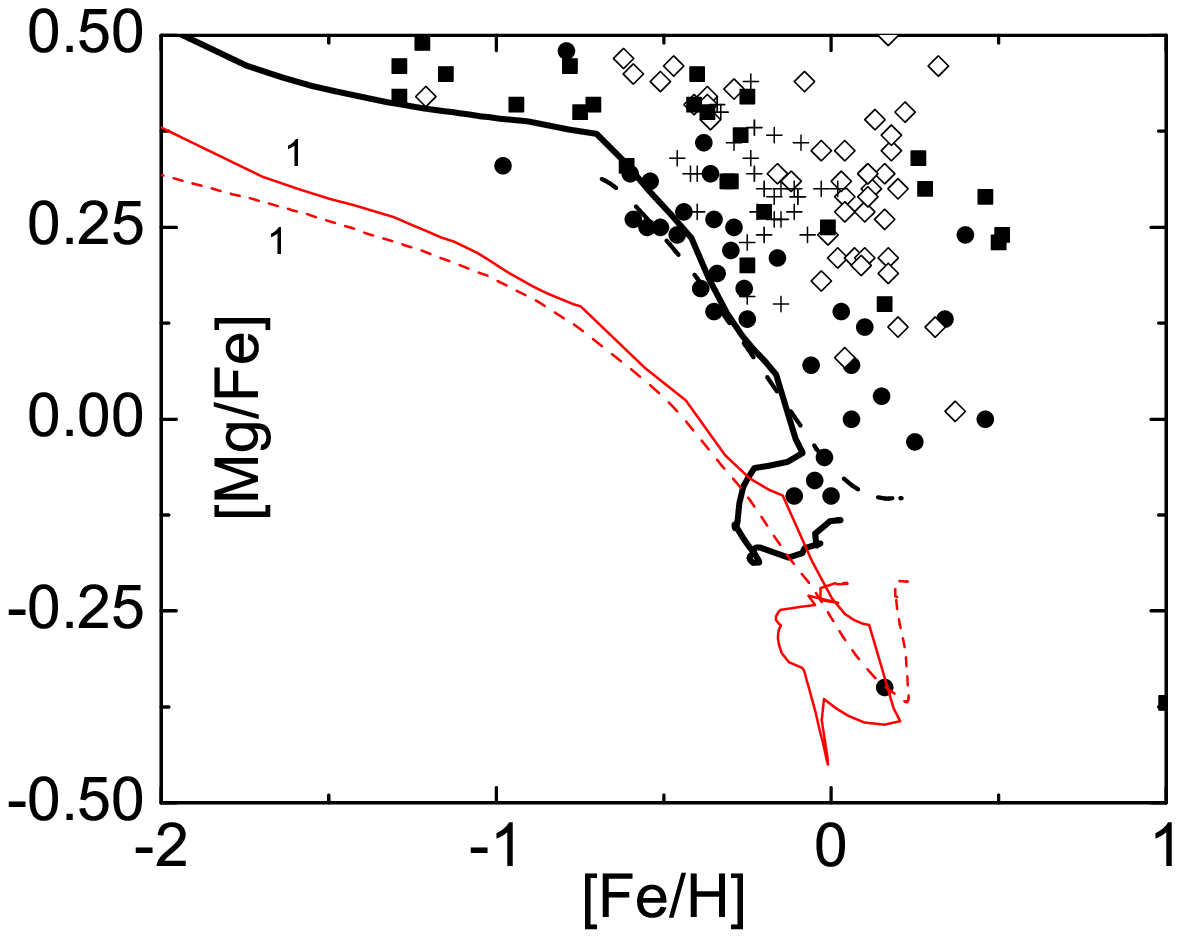,width=65truemm,angle=0,clip=}}
\vspace{-5mm}
\captionc{4}{[Si/Fe] and [Mg/Fe] as a function of [Fe/H] for bulge stars. Symbols are 
             as in Figure~3.}}
\vspace{5mm}

Other $\alpha$-elements also show a similar behaviour with metallicity as shown in Figure~4 
for the ratios [Si/Fe] and [Mg/Fe]. The symbols are the same as in Figure~3 and the same 
comments regarding our models apply here.  The agreement is reasonable, taking into account 
that the intrinsic dispersion of the data is much higher. Although these ratios are not
as well determined as in the case of oxygen, it is interesting to note that the multizone
models show a comparatively better fit to the data than the single zone models.

\subsectionb{5.4}{Abundance correlations in PNe: O, Ne, S, and Ar}

Distance independent correlations of chemical abundances of elements that are not 
manufactured in the PNe progenitor stars also provide interesting constraints for chemical 
evolution models, a procedure already successfully used in the galactic disk. Figure~5
shows correlations with oxygen of the Ne, S, and Ar abundances. It can be seen that both
classes of models are able to explain reasonably the observed correlations, taking into
account the average uncertainties of the abundance determinations, which may reach
about 0.2 to 0.3 dex.

\vspace{5mm}
\vbox{
\centerline{\psfig{figure=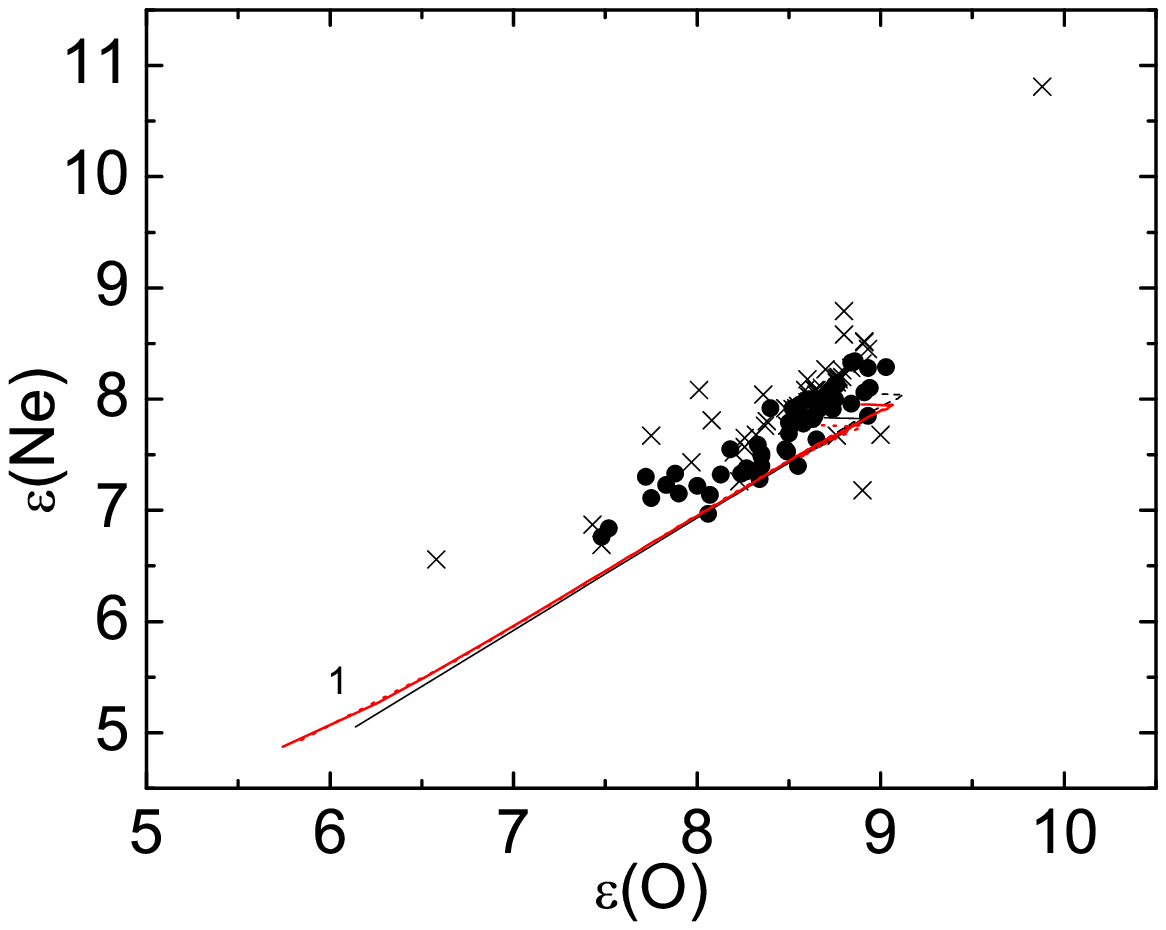,width=45truemm,angle=0,clip=}
\psfig{figure=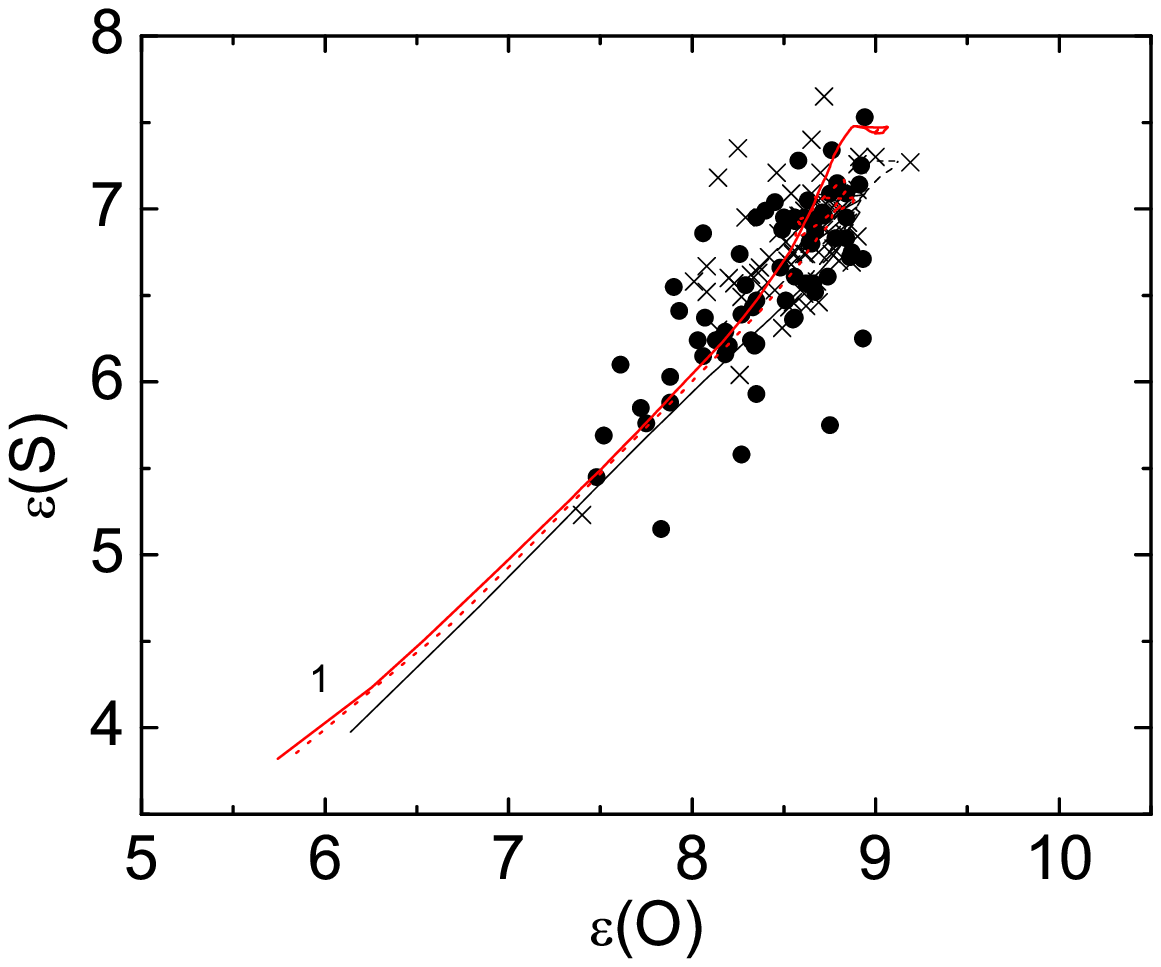,width=45truemm,angle=0,clip=}
\psfig{figure=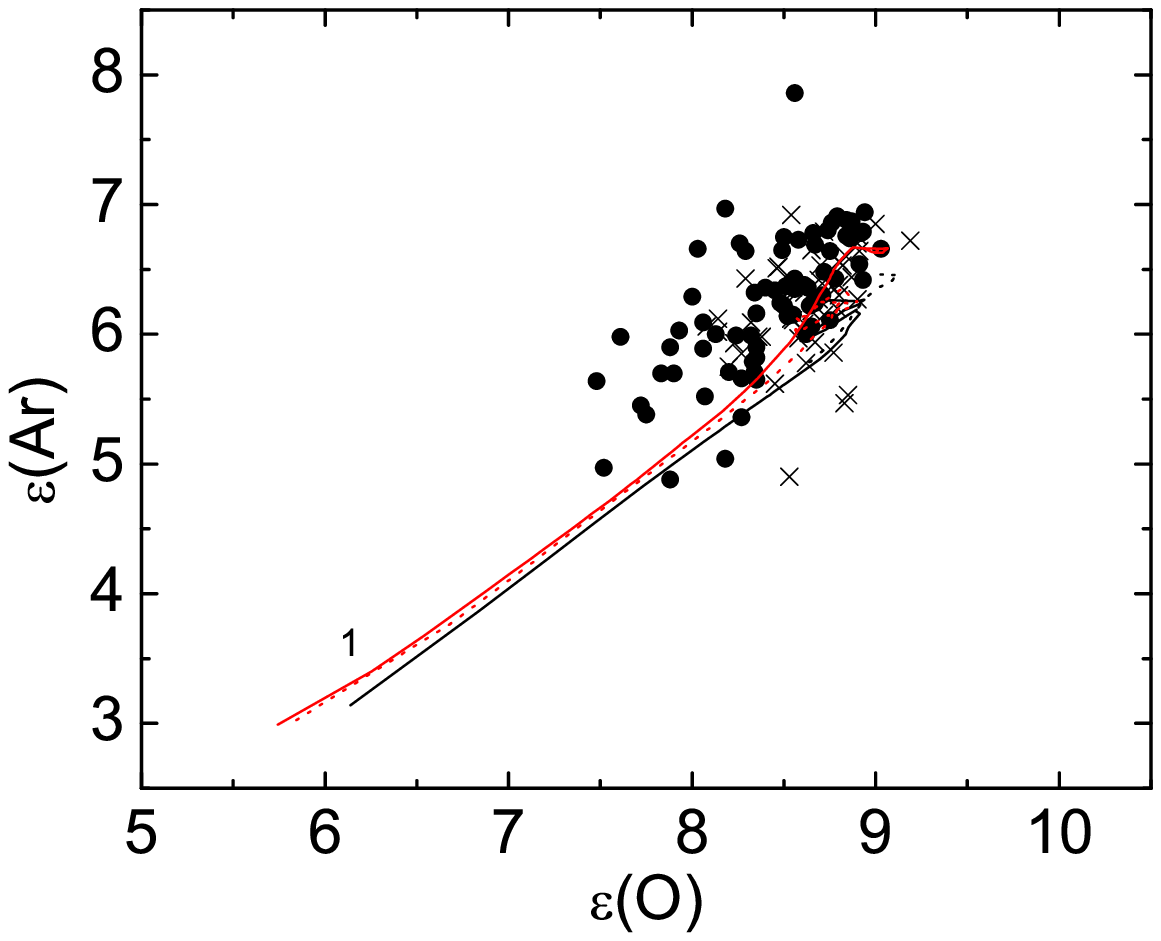,width=45truemm,angle=0,clip=}}
\vspace{-5mm}
\captionc{5}{Distance independent correlations of the abundance of Ne, S, and Ar 
             compared to model predictions. Models symbols are as in Figure~1.}}
\vspace{5mm}

\subsectionb{5.5}{Nitrogen abundances}
Nitrogen is an interesting element, in the sense that its abundance can be accurately 
determined in PNe, in the same way as oxygen. The main difference is that part of the 
observed nebular abundances is probably due to the dredge-up episodes occurring in the 
PNe progenitor stars, which must be taken into account when interpreting N abundances.
The N enhancements are modest, and are especially important in the so-called Type I PNe 
(Peimbert 1978), which comprise a rather small fraction of the observed nebulae. This is 
recognized for example in the recent models for the bulge by Ballero et al. (2007b), 
where a sample of bulge PNe was taken into account. Moreover, it has long been known 
from disk PNe that the N-rich objects have usually $\epsilon({\rm N}) = \log({\rm N/H})+12 > 
8.0$ (cf. Fa\'undez-Abans and Maciel 1987), and these objects belong to the high mass end 
of the intermediate mass stars that originate the PNe, which comprises a relatively small 
fraction of all PNe progenitor stars. Most of these objects have N-enhancements of a few 
tenths in the logarithmic scale, while the average N abundance of PNe is about 
$\epsilon({\rm N}) \simeq 8.0$. Therefore, it is expected that  most PNe in the bulge have 
approximately normal N abundances, a result fully supported by our previous data (Cuisinier 
et al. 2000, Escudero and Costa 2001, Escudero et al. 2004), so that N could be used -- 
although cautiously -- in the comparison with theoretical models, an approach taken by 
Ballero et al. (2007b).

\vspace{3mm}
\vbox{
\centerline{\psfig{figure=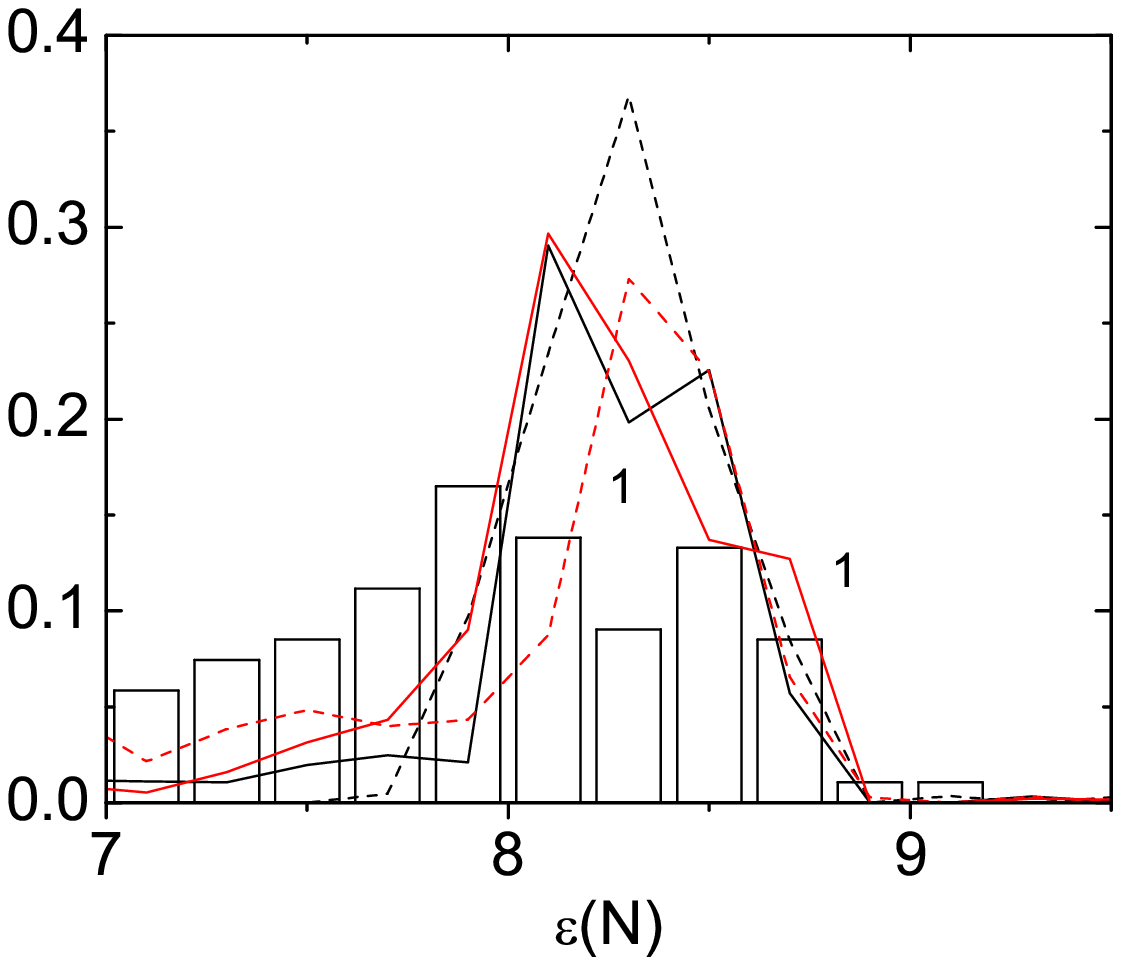,width=65truemm,angle=0,clip=}
\psfig{figure=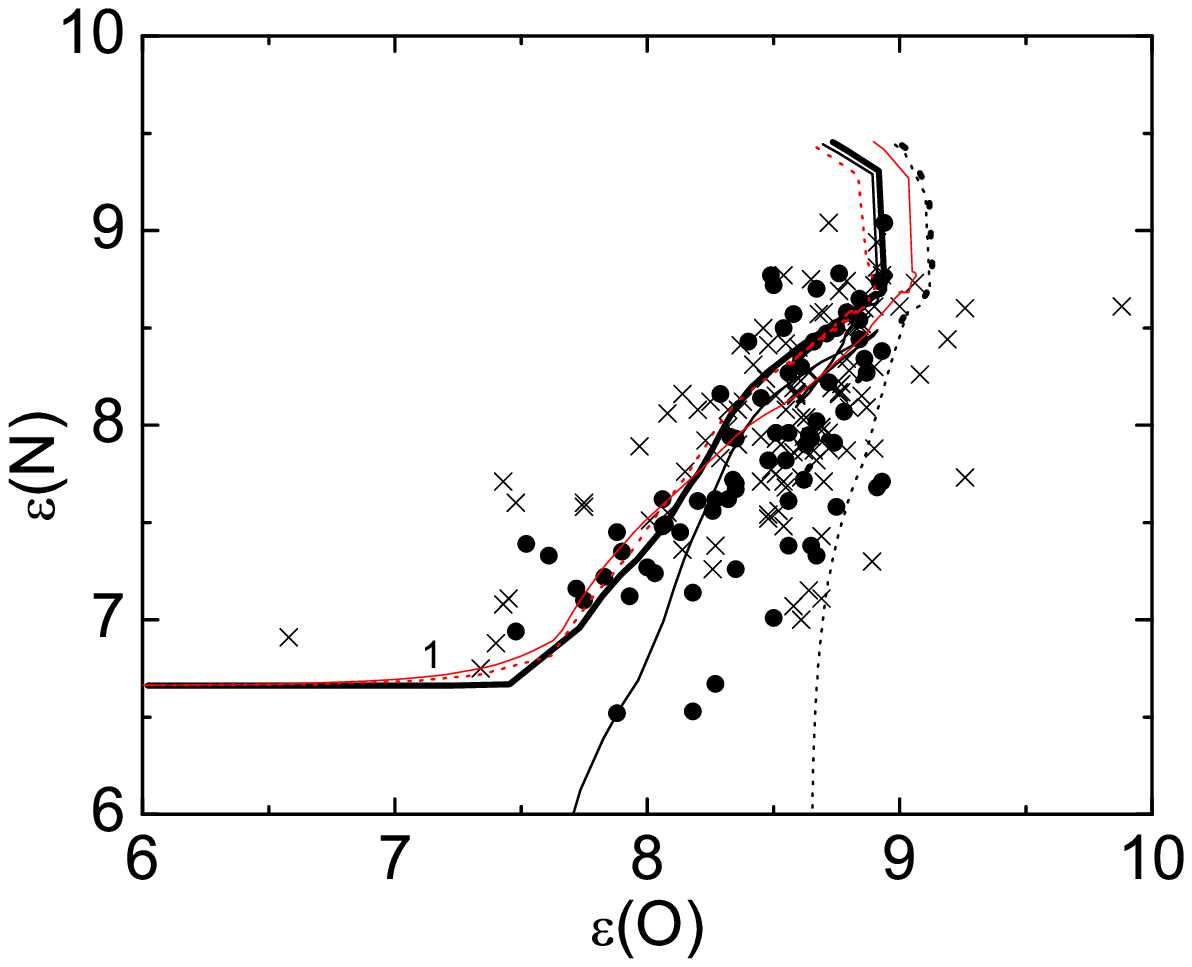,width=65truemm,angle=0,clip=}}
\vspace{-5mm}
\captionc{6}{(Left) Abundance distribution of nitrogen compared to model predictions.
             Symbols are the same as in Figure~1.}
\captionc{7}{(Right) The same as Figure~5 for nitrogen. Here the multizone model is displayed as 
             black thick lines for the van den Hoek \& Groenewegen (1997) yields, 
             black  thin  lines adopting no yields for N and O for stars with masses 
             smaller than one solar mass and black dotted lines for zone 1.}}
\vspace{3mm}

In Figure~6 we show the observed N abundance distribution along with the theoretical
models. Symbols are the same as in Figure~1. We confirm that N abundances peak around
$\epsilon(N) \simeq 8$, within the uncertainties, comparable to the disk distribution. 
The distribution falls abruptly towards higher N abundances, confirming that few objects 
are strongly N-enhanced and that there is no strong evidence for a recent star formation 
in the bulge, which would produce many young, massive stars with a high N-enhancement. The 
nitrogen abundances derived from both models are similar, the multizone models producing 
a slightly better fit to the data in spite of presenting a narrower distribution with 
respect to the observational data, and taking into account the probable incompleteness
of the observational sample.

In Figure~7 we show the distance-independent correlation of $\epsilon$(N) as a function
of $\epsilon$(O) for bulge PNe, which may be compared with Figure~5. Here the black thick 
lines represent results derived using van den Hoek \& Groenewegen (1997) yields and 
thin black lines represent results  derived adopting no yields for nitrogen and oxygen  
for stars with masses smaller than one solar mass. In these plots the different symbols 
refer to PNe from different data samples (see Escudero et al. 2004 for details).

From Figure~7 it can be seen that both models are able to reproduce reasonably well most 
of the data. It can be seen that there is a large scatter in the oxygen abundances for 
objects with low nitrogen abundances. This scatter can be reproduced assuming that winds 
produced by SNe of types II and Ia are responsible for the loss of elements synthesized 
by the SNe, leading to a chemical enrichment of heavier elements such as oxygen and iron 
relative to lighter elements such as nitrogen and helium.  There is also a large number 
of objects with high oxygen abundances, but with low nitrogen. All one-zone models or 
those with only primordial gas infall cannot describe this population, which may indicate 
that these objects originate from an already oxygen-enriched medium. Since it is produced 
mainly by SNII, this result suggests that this population was formed in an epoch where 
the gas was already enriched by material ejected by stars with short lifetimes in a medium 
not yet enriched with elements produced by longer lifetime stars, like nitrogen.

\sectionb{6} {DISCUSSION}

From the results presented in the previous section, there are several evidences in
favour of a multizone, double-infall model for the evolution of the galactic bulge,
in comparison to a less complex model.  In fact, double-infall models have been successfully 
used to build chemical evolution models for the galactic disk. For example, in the 
scenario devised by Chiappini et al. (1997), the first infall  (with $1.24 \times 
10^{10} M_{\odot}$ and 0.1 Gyr timescale) would be responsible for the formation of 
the old stellar population detected throughout the bulge, and the second episode 
would represent the gas infall that formed the disk (with $2.26 \times 10^9 M_{\odot}$ 
and 2 Gyr timescale), beginning 2 Gyr after the first one. In this model 
the formation of the  bulge is included in the first infall episode forming the halo, 
and the second infall episode is applied to the disk, while our model assumes two 
infall episodes to explain the observed properties of the bulge.  Any chemical 
evolution model for the bulge must have a large gas infall episode at the beginning 
of its formation, in order to reproduce the old population present in that region. 

An important observational constraint to the hypothesis of a second infall episode is 
the presence of an intermediate mass population in the bulge, seen either as PNe or 
stars (van Loon et al. 2003), since these objects would result directly from the second 
infall. Observational evidences of a central bar (Bissantz \& Gerhard 2002) also favour 
the hypothesis of a second infall. Therefore, we developed also a double-infall model to 
better reproduce the observed chemical abundance pattern. In this model we assume 
that 2\% of the mass for the objects between 3 and 16 $M_{\odot}$ generate binary systems 
that eventually will become SNIa.  Yoshii et al. (1996) derived values between 2\% and 
2.5\% when assuming Scalo (1986) IMF and values between 5\% and 5.5\% adopting Salpeter's 
IMF. Adopting Kroupa's (2002) IMF, which is very similar to that by Scalo, a value of 
2\% is obtained, consistent with disk data. 

The galactic bulge had its first star formation episode resulting from a collapse of 
primordial gas. This episode is a consensus among the recent star formation models. The 
need of this initial scenario is based not only on results from chemical evolution models 
but mainly on observational data. Among these is the presence of a large number of old 
objects, which requires the existence of an extensive star formation at the beginning of 
the galactic evolution. Also, the wide metallicity distribution found in stars and PNe
has to be taken into account. According to the present model, a fast gas collapse is 
more efficient to form a wide range of metallicities than a slow infall process. A third 
observational constraint is the presence of old, metal rich  stars. An abrupt collapse 
is able to rapidly enrich the gas, generating both metal-poor and metal-rich stars.

The main characteristics of the first infall is a large mass loss to the outer regions
such as the halo, disk or even out of the galaxy, produced by SNII/Ia. This loss of metals 
is essential to reproduce the observed abundance distributions of PNe. This is 
also a consensus among the evolutionary models for the bulge. Without this process, stellar 
abundances would be higher than observed. From our results, the fraction of ejected gas
cannot be defined exactly, due to multiple assumptions in the model input. However, it is 
clear from our results that the multizone model reproduces better the observational 
constraints. To define this fraction accurately, as well as its time dependence, more 
accurate observational data and more realistic hydrodynamical models for the central 
region are required.

The fate of the ejected material is not clear. Samland et al. (1997) suggest that this 
material was ejected to the halo and eventually fell onto the disk. These authors are able 
to reproduce different chemical properties of the interstellar medium and the disk. 
However, one of their conclusions is that the abundance gradient only begun after 6 Gyr, 
in contradiction to the results from PNe and open clusters (cf. Maciel et al. (2003, 2005, 
2006, 2007; Friel et al. 2002), that show the existence of an expressive gradient at that epoch. 
In the present work, we adopt radial fluxes produced by SN, whose displacement are 
restricted to the adjacent zone at 1.5 kpc. With this assumption we were able to explain 
the presence of oxygen-rich and nitrogen-poor planetary nebulae in the bulge.
 
\sectionb{6} {CONCLUSIONS}

We developed three classes of models to reproduce the abundances of the PNe population of the 
galactic bulge, representing the chemical evolution of its intermediate mass population, as
well as recent data on bulge stars. An effort was made to increase the amount of observational
constraints to be explained by the models, so that their reliability is enhanced even though
the agreement with the observations might not be perfect. The model results were compared 
to recent observational data of PNe and stars, leading to the following conclusions:
(i) Most of the abundances can be reproduced assuming a fast initial collapse with a high 
wind rate (ii) Some peculiarities found in the abundances of PNe and bulge stars require 
the existence of a second infall of material previously enriched by SNII ejecta, and 
(iii) Abundance ratios from stars ($\alpha$/Fe) suggest that, at the beginning of the 
bulge formation, the IMF was steeper. The best way to describe it is to assume Salpeter's 
IMF for the initial 0.6 Gyr and Kroupa's for the rest of the evolution. 

\vspace{5mm}

ACKNOWLEDGMENTS. This work was partly supported by the Brazilian agencies FAPESP and CPNq.      
A.V.E acknowledges FAPESP for his graduate fellowship (Proc. 00/12609-0).

\vspace{5mm}

\References

\refb
Amaral~L.~H., Ortiz~R., L\'epine~J.~R.~D., Maciel~W.~J. 1996, MNRAS, 281, 339

\refb
Ballero~S.~K., Matteucci~F., Chiappini~C. 2006, New Astr. 11, 306

\refb
Ballero~S.~K., Kroupa~P., Matteucci~F. 2007a, A\&A 467, 117

\refb
Ballero~S.~K., Matteucci,~F., Origlia,~L., Rich~R.~M. 2007b, A\&A 467, 123

\refb
Barbuy~B., Grenon~M. 1990, ESO/CTIO Workshop on Bulges of Galaxies, p.83

\refb
Beaulieu~S.~F., Freeman~K.~C., Kalnajs~A.~J., Saha~P. 2000, AJ 120, 855

\refb
Bissantz~N., Gerhard~O. 2002, MNRAS 330, 591

\refb
Boselli~A. 1994, A\&A 292, 1

\refb
Boselli~A., Gavazzi~G., Lequeux~J., Buat~V., Casoli~F., et al. 
    1995, A\&A 300, L13

\refb
Buat~V., Deharveng~J.~M., Donas~J. 1989, A\&A 223, 42

\refb
Buat~V. 1992, A\&A 264, 444


\refb
Cavichia~O., Costa~R.~D.~D., Maciel~W.~J. 2008, (in preparation)

\refb
Chiappini~C., Matteucci~F., Gratton~R. 1997, ApJ 477, 765

\refb
Chiosi~C. 1980, A\&A 83, 206

\refb
Cuisinier~F., Maciel~W.~J., K\"oppen~J., Acker~A., Stenholm~B. 2000,  A\&A 353, 543

\refb
Cunha~K., Smith~V.~V. 2006, ApJ 651, 491

\refb
Deharveng~J.~M., Sasseen~T.~P., Buat~V., Bowyer~S., Lampton~M., Wu~X. 1994, A\&A 289, 715

\refb
Escudero~A.~V., Costa~R.~D.~D. 2001, A\&A 380, 300

\refb
Escudero~A.~V, Costa~R.~D.~D., Maciel~W.~J. 2004, A\&A 414, 211

\refb
Fa\'undez-Abans~M., Maciel~W.~J. 1987, A\&A 183, 324

\refb
Ferreras~I., Wyse~R.~F.~G., Silk~J. 2003, MNRAS 345, 1381

\refb
Ferrini~F., Matteucci~F., Pardi~C., Penco~U. 1992, ApJ 387, 138

\refb
Friel~E.~D., Janes~K.~A., Tavarez~M., Scott~J., Katsanis~R., Lotz~J., 
    Hong~L., Miller~N. 2002, AJ, 124, 2693


\refb
Fulbright~J.~P., McWilliam~A., Rich~R.~M. 2006, ApJ 636, 821

\refb
Fulbright~J.~P., McWilliam~A., Rich~R.~M. 2007, ApJ 661, 1152



\refb
Kennicutt~R.~C. 1989 ApJ 344, 685

\refb
Kennicutt~R.~C. 1998a, ApJ 498, 541

\refb
Kennicutt~R.~C. 1998b, ARA\&A 36, 189

\refb
Kroupa~P. 2002, Science 295, 82


\refb
Lecureur~A., Hill~V., Zoccali~M., Barbuy~B., G\'omez~A., Minniti~D., 
    Ortolani~S., Renzini~A. 2007, A\&A 465, 799


\refb
Maciel~W.~J., Costa~R.~D.~D., Uchida~M.~M.~M. 2003, A\&A 397, 667

\refb
Maciel~W.~J., Lago~L.~G., Costa~R.~D.~D. 2005, A\&A 433, 127

\refb
Maciel~W.~J., Lago~L.~G., Costa~R.~D.~D. 2006, A\&A 453, 587

\refb
Maciel~W.~J., Quireza~C., Costa~R.~D.~D. 2007, A\&A 463, L13




\refb
Matteucci~F. 2001, The Chemical evolution of the Galaxy, Kluwer

\refb
Matteucci~F., Brocato~E. 1990, ApJ 365, 539

\refb
Matteucci~F., Greggio~L. 1986, A\&A 154, 279

\refb
Matteucci~F., Panagia~N., Pipino~A., Mannucci~F., Recchi~S., Della~Valle~M., 
    2006, MNRAS 372, 265

\refb
McWilliam~A., Matteucci~F., Ballero~S., Rich~R.~M., Fulbright~J.~P., Cescutti~G. 2008, AJ 136, 367





\refb
Moll\'a~M., Ferrini~F., Gozzi~G. 2000, MNRAS 316, 345


\refb
Padoan~P., Nordlund~A.~P., Jones~B.~J. 1997, MNRAS 288, 145

\refb
Peimbert~M. 1978, IAU Symp. 76, ed. Y. Terzian, Reidel, p. 215

\refb
Perinotto~M., Bencini~C.~G., Pasquali~A., Manchado~A.,
    Rodriguez Espinosa~J.~M., Stanga~R. 1999, A\&A 347, 967

\refb
Pomp\'eia~L., Barbuy~B., Grenon~M. 2003, ApJ 592, 1173


\refb
Rana~N.~C. 1991, ARA\&A 29, 129




\refb
Rich~R.~M., Origlia~L. 2005, ApJ 634, 1293

\refb
Rich~R.~M., Origlia~L., Valenti~E. 2007, ApJ 665, L119

\refb
Samland~M., Hensler~G., Theis~Ch. 1997, ApJ 476, 544

\refb
Scalo~J.~M. 1986, Fund. Cosmic Phys. 11, 1

\refb
Schmidt~M. 1959, ApJ 129, 243

\refb
Silk~J. 1995, ApJ 438, L41





\refb
Tsujimoto~T., Nomoto~K., Yoshii~Y., Hashimoto~M., Yanagida~S.,  
    Thielemann~F.~K. 1995, MNRAS 277, 945


\refb
van den Hoek~L.~B., Groenewegen~M.~A.~T. 1997, A\&AS 123, 305

\refb
van Loon~J.~Th., Gilmore~G.~F., Omont~A. et al. 2003 MNRAS 338, 857



\refb
Yoshii~Y., Tsujimoto~T., Nomoto~K.~I. 1996, ApJ 462, 266


\refb
Zoccali~M., Hill,~V., Lecureur,~A., Barbuy,~B., Renzini~A., Minniti, D., G\'omez,~A., Ortolani~S. 
    2008, A\&A 486, 177

\refb
Zoccali~M., Lecureur~A., Barbuy~B., Hill~V., Renzini~A., 
    Minniti~D., Momany~Y., G\'omez~A., Ortolani~S.  2006, A\&A 457, L1

\refb
Zoccali~M., Renzini~A., Ortolani~S., Greggio~L., Saviane~I., 
    Cassisi~S., Rejkuba~M., Barbuy~B., Rich~R.~M., Bica~E. 2003, A\&A 399, 931

\end{document}